\documentclass[sigconf]{acmart}

\renewcommand\footnotetextcopyrightpermission[1]{} 
\makeatletter
\renewcommand\@formatdoi[1]{\ignorespaces}
\makeatother

\acmConference[SIGMETRICS 2020]{SIGMETRICS 2020}{June 08--12, 2018}{Boston, MA}

\usepackage{amsmath,bm,calc,amssymb}

\let\oldhat\hat

\renewcommand{\hat}[1]{\oldhat{\boldsymbol{\mathbf{#1}}}}

\usepackage{textcomp}
\usepackage{color}

\usepackage[ruled,noline]{algorithm2e}

\usepackage{subcaption}
\usepackage{epsf}
\newcommand{\epsin}[2]%
  {\setlength{\epsfxsize}{#2\hsize}\centerline{\epsfbox{#1}}}

\usepackage{stfloats}

\usepackage{tabularx}
\usepackage{multirow}

\usepackage{paralist}

\usepackage{listings}
\usepackage{color}

\definecolor{dkgreen}{rgb}{0,0.6,0}
\definecolor{gray}{rgb}{0.5,0.5,0.5}
\definecolor{mauve}{rgb}{0.58,0,0.82}

\usepackage{graphicx}

\begin{document}

\title[]{Fast Dimensional Analysis for Root Cause Investigation in a Large-Scale Service Environment}

\author{Fred Lin}
\affiliation{%
  \institution{Facebook, Inc.}}
\email{fanlin@fb.com}

\author{Keyur Muzumdar}
\affiliation{%
  \institution{Facebook, Inc.}}
\email{kmuzumdar@fb.com}

\author{Nikolay Pavlovich Laptev}
\affiliation{%
  \institution{Facebook, Inc.}}
\email{nlaptev@fb.com}

\author{Mihai-Valentin Curelea}
\affiliation{%
  \institution{Facebook, Inc.}}
\email{mihaic@fb.com}

\author{Seunghak Lee}
\affiliation{%
  \institution{Facebook, Inc.}}
\email{seunghak@fb.com}

\author{Sriram Sankar}
\affiliation{%
  \institution{Facebook, Inc.}}
\email{sriramsankar@fb.com}

\begin{abstract}

Root cause analysis in a large-scale production environment is challenging due to the complexity of services running across global data centers. Due to the distributed nature of a large-scale system, the various hardware, software, and tooling logs are often maintained separately, making it difficult to review the logs jointly for understanding production issues. Another challenge in reviewing the logs for identifying issues is the scale - there could easily be millions of entities, each described by hundreds of features. In this paper we present a \textit{fast dimensional analysis} framework that automates the root cause analysis on structured logs with improved scalability.

We first explore item-sets, \textit{i.e.} combinations of feature values, that could identify groups of samples with sufficient \textit{support} for the target failures using the Apriori algorithm and a subsequent improvement, FP-Growth. These algorithms were designed for frequent item-set mining and association rule learning over transactional databases. After applying them on structured logs, we select the item-sets that are most unique to the target failures based on \textit{lift}. We propose pre-processing steps with the use of a large-scale real-time database and post-processing techniques and parallelism to further speed up the analysis and improve interpretability, and demonstrate that such optimization is necessary for handling large-scale production datasets. We have successfully rolled out this approach for root cause investigation purposes in a large-scale infrastructure. We also present the setup and results from multiple production use cases in this paper.

\end{abstract}

\maketitle

\section{Introduction}
\label{s:introduction}

Companies running Internet services have been investing in autonomous systems for managing large scale services, for better efficiency and scalability~\cite{Greenberg09.1}. As some of the Internet services have become utilities that the public relies on for transportation, communication, disaster response, \textit{etc.}, the reliability of the infrastructure is now emphasized more than before. There are various logs that these systems record and act upon. The logs record events and configurations about the hardware, the services, and the automated tooling, which are important in measuring the performance of the system and tracing specific issues. Given the distributed nature and the scale of a modern service environment, it is challenging to find and monitor patterns from the logs, because of the scale and the complexity of the logs - each component in the system could record millions of entities that are described by hundreds of features. An automated RCA (Root Cause Analysis) tool is therefore needed for analyzing the logs at scale and finding strong associations to specific failure modes.

Traditional supervised machine learning methods such as logistic regression are often not interpretable and require manual feature engineering, making them impractical for this problem. Castelluccio \textit{et al.} proposed to use STUCCO, a tree-based algorithm for contrast set mining~\cite{Bay99.1} for analyzing software crash reports~\cite{Castelluccio17.1}. However, the pruning process in STUCCO could potentially drop important associations, as illustrated in Section~\ref{ss:proposed_framework}.

In this paper, we explain how we modified the classical frequent pattern mining approach, \textit{Apriori}~\cite{Agrawal93.1}, to handle our root cause investigation use case at scale. While Apriori has been an important algorithm historically, it suffers from a number of inefficiencies such as its runtime and the expensive candidate generation process. The time and space complexity of the algorithm are exponential $O(2^D)$ where $D$ is the total number of items, \textit{i.e.} feature values, and therefore it is practical only for datasets that can fit in memory. Furthermore, the candidate generation process creates a large number of item-sets, \textit{i.e.} combinations of feature values, and scans the dataset multiple times leading to further performance loss. For these reasons, \textit{FP-Growth} has been introduced which significantly improves on Apriori’s efficiency.

FP-Growth is a more efficient algorithm for frequent item-set generation~\cite{Han00.1}. Using the divide-and-conquer strategy and a special frequent item-set data structure called \textit{FP-Tree}, FP-Growth skips the candidate generation process entirely, making the algorithm more scalable and applicable to datasets that cannot fit in memory. As we show in the experimental results in Section~\ref{s:experimental_result}, FP-Growth can be $50\%$ faster than a parallelized Apriori implementation when the number of item-sets is large.

While FP-Growth is significantly more efficient than Apriori, some production datasets in large-scale service environments are still too large for FP-Growth to mine all the item-sets quickly for time-sensitive debugging. The huge amount of data could also become a blocker for memory IO or the transfer between the database and local machines that run FP-Growth. To further speed up the analysis, we use Scuba~\cite{Abraham13.1}, a scalable in-memory database where many logs are stored and accessed in real-time. As many recorded events in the production logs are identical except for the unique identifiers such as timestamps and job IDs, we \textit{pre-aggregate} the events using Scuba's infrastructure before querying them for the root cause analysis. The pre-aggregation step saves runtime and memory usage significantly, and is necessary for enabling automatic RCA on production datasets at this scale.

The framework lets users specify irrelevant features, \textit{i.e.} columns, in the structured log to be excluded for avoiding unnecessary operations, thereby optimizing the performance. Users can also specify the \textit{support} and \textit{lift} of the analysis for achieving the desired tradeoff between the granularity of the analysis and the runtime. For example, a faster and less granular result is needed for mission critical issues that need to remediated immediately; and more thorough results from a slower run are useful for long-term analyses that are less sensitive to the runtime. Parallelism and automatic filtering of irrelevant columns are also features for achieving better efficiency, which we discuss in Section~\ref{ss:performance_optimization}.

With the above optimizations, we have productionized a \textit{fast dimensional analysis} framework for the structured logs in a large-scale infrastructure. The fast dimensional analysis framework has found various association rules based on structured logs in different applications, where the association rules reveal hidden production issues such as anomalous behaviors in specific hardware and software configurations, problematic kernel versions leading to failures in auto-remediations, and abnormal tier configurations that led to an unexpectedly high number of exceptions in services.

The rest of the paper is organized as follows: We discuss the requirements of a large-scale service environment, the advantage of logging in a structured format, the typical RCA flow in Section~\ref{s:rca_in_system}. We illustrate the proposed framework in Section~\ref{s:fda}. We demonstrate the experimental results in Section~\ref{s:experimental_result}, and the applications on large-scale production logs in Section~\ref{s:use_case}. Section~\ref{s:conclusion} concludes the paper with a discussion on future work.

\section{Root Cause Analysis in a Large-Scale Service Environment}
\label{s:rca_in_system}

\thispagestyle{plain} \subsection{Architecture of a Large-Scale Service Environment}
\label{ss:architecture_of_system}

Large scale service companies like Google, Microsoft, and Facebook have been investing in data centers to serve globally distributed customers. These infrastructures typically have higher server-to-administrator ratio and fault tolerance as a result of the automation that is required for running the services at scale, and the flexibility to \textit{scale out} over a large number of low-cost hardwares instead of \textit{scaling up} over a smaller set of costly machines~\cite{Greenberg09.1}. Two important parts for keeping such large-scale systems at high utilization and availability are resource scheduling and failure recovery.

Resource scheduling mainly focuses on optimizing the utilization over a large set of heterogeneous machines with sufficient fault tolerance. Various designs of resource scheduling have been well-documented in literature, such as Borg from Google~\cite{Verma15.1}, Apollo from Microsoft~\cite{Boutin14.1}, Tupperware from Facebook~\cite{Yu19.1}, Fuxi from Alibaba~\cite{Zhang14.1}, Apache Mesos~\cite{Hindman11.1} and YARN~\cite{Vavilapalli13.1}.

The ultimate goal for a failure recovery system is to maintain the fleet of machines at high availability for serving applications. Timely failure detection and root cause analysis (RCA), fast and effective remediation, and proper spare part planning are some of the keys for running the machines at high availability. While physical repairs still need to be carried out by field engineers, most parts in a large-scale failure recovery system have been fully automated to meet the requirements for high availability. Examples of the failure handling systems are Autopilot from Microsoft~\cite{Isard07.1} and FBAR from Facebook~\cite{Lin18.1}.

\thispagestyle{plain} \subsection{Logs in a Large-Scale System}
\label{ss:logs_in_system}

Proper logging is key to effectively optimizing and maintaining a large-scale system. In a service environment composed of heterogeneous systems, logs come from three major sources:

\begin{itemize}
\item \textbf{Software} - The logs populated from the services running on the servers are critical for debugging job failures. Job queue times and execution times are also essential for optimizing the scheduling system. Typically program developers have full control in how and where the events should be logged. Sometimes a program failure needs to be investigated together with the kernel messages reported on the server, \textit{e.g.} out of memory or kernel panic.
\item \textbf{Hardware} - Hardware telemetries such as temperature, humidity, and hard drive or fan spinning speed, are collected through sensors in and around the machines. Hardware failures are logged on the server, \textit{e.g.} System Event Log (SEL) and kernel messages (dmesg). The hardware and firmware configurations of the machine are also critical in debugging hardware issues, \textit{e.g.} the version of the kernel and the firmwares running on different components. The messages on the servers need to be polled at an appropriate frequency and granularity that strikes a balance between the performance overhead on the servers and our ability to detect the failures timely and accurately. 
\item \textbf{Tooling} - As most of the parts in the large-scale system are automated, it is important to monitor the tools that orchestrate the operations. Schedulers would log the resource allocations and job distribution results. Failure recovery systems would log the failure signals and the remediation status. Historical tooling logs are important for analyzing the tooling efficiency. 
\end{itemize}


For root cause analysis in real-time, the logs are pushed to Scuba, a “fast, scalable, distributed, in-memory database.”~\cite{Abraham13.1} Keeping the data in-memory, Scuba tables typically have shorter retention. For long-term analytics, logs are archived in disk-based systems such as HDFS~\cite{Borthakur19.1}, which can be queried by Hive~\cite{Thusoo10.1} and Presto~\cite{Traverso13.1}. Some of the more detailed operational data can be fetched from the back-end MySQL databases~\cite{MySQL_FB} to enrich the dataset for the analysis. The quality of the logs has fundamental impacts on the information we can extract. We will discuss the advantages of structured logging in Section~\ref{ss:structured_data}.

\thispagestyle{plain} \subsection{Prior Root Cause Analysis Work}
\label{ss:prior_work}

Root cause analysis (RCA) is a systematic process for identifying the \textit{root causes} of specific events, \textit{e.g.} system failures. RCA helps pinpoint contributing factors to a problem or to an event. For example, RCA may involve identifying a specific combination of hardware and software configurations that are highly correlated to unsuccessful server reboots (discussed in Section~\ref{ss:anomalous_hw_and_sw}), and identifying a set of characteristics of a software job that are correlated to some types of job exceptions (discussed in Section~\ref{ss:failed_auto_remediation}).

During an incident in a large-scale system, the oncall engineers typically investigate the underlying reason for a system failure by exploring the relevant datasets. These datasets are comprised of tables with numerous columns and rows, and often the oncall engineers would try to find aggregations of the rows by the column values and correlate them with the error rates. However, a naive aggregation scales poorly due to the significant amount of the rows and distinct values in the columns, which result in a huge amount of groups to be examined.

For automating RCA, the STUCCO algorithm has been used for contrast set mining \cite{Bay99.1,Bay01.1,Castelluccio17.1}. Suriadi \textit{et al.} \cite{Suriadi12.1} demonstrated an RCA approach using decision tree-based classifications from the WEKA package~\cite{Witten17.1}, as well as enriching the dataset with additional features. The traditional STUCCO algorithm, however, can miss important associations if one of the items does not meet the the pruning threshold on the $\chi^2$ value, as explained in Section~\ref{ss:proposed_framework}. Decision tree-based approaches, while providing the visibility in how the features are used to construct the nodes, become harder to tune and interpret as the number of trees grows. To ensure we capture the associations that are relatively small in population yet strongly correlated to our target, we choose to explore all association rules first with additional filtering based on support and lift as post-processing, as illustrated in Section~\ref{ss:interpretability_optimization}.

In association rule mining, FP-Growth~\cite{Han00.1} has become the common approach as the classical Apriori algorithm~\cite{Agrawal93.1} suffers from its high complexity. The state-of-the-art approaches for log-based root cause analysis found in literature often have small-data experiments, do not handle redundant item-sets with pre-/post-processing and fail to find root causes with small error rates relative to successes. For example, while the authors in \cite{Zhang18.1} provide a pre-processing step for pre-computing the association matrix for speeding up association rule mining, it lacks the study of large scale applicability and the filtering of small error classes (see Figure \ref{f:assoc_rule_vs_min_lift} and Section \ref{ss:interpretability_improvement} for how the proposed fast dimensional analysis framework addresses these points). Similarly, the authors in \cite{Arifin16.1} provide a hybrid approach for spam detection using a Naive Bayes model with FP-Growth to achieve better spam-detection accuracy but with a decreased interpretability because only the prediction is provided and not the root-cause. In our use case, however, having an explainable model is critical (see Section \ref{ss:interpretability_improvement}) and we are biased away from compromises in interpretability.

An FP-Growth implementation on Spark platform was proposed in~\cite{Liu16.1}, and an extension for FP-Growth to handle negative association rules \cite{Wong05.1} was proposed in \cite{Wang17.1}. For better interpretability, Bittmann~\textit{et al.} proposed to remove ubiquitous items from the item-sets using \textit{lift}~\cite{Bittmann18.1}, whereas Liu~\textit{et al.} used lift to further prune the mined association rules from FP-Growth~\cite{Liu16.1}.

For sustaining the large-scale service environment at high availability all the time, we need an analysis framework that can handle large-scale production datasets, \textit{e.g.} billions of entries per day, and generate result in near real-time, \textit{e.g.} seconds to minutes. In this paper we propose a framework that pre-aggregates data to reduce data size by $>500X$ using an in-memory database~\cite{Abraham13.1}, mines frequent item-sets using a modified version of FP-Growth algorithm, and filters the identified frequent item-sets using support and lift for better interpretability. We validate the framework on multiple large-scale production datasets from a service environment, whereas the related papers mostly demonstrate the results using relatively small synthetic datasets. We will illustrate the details of the framework in Section~\ref{s:fda} and compare the framework with the above-mentioned methods in Section~\ref{s:experimental_result}.

\section{Fast Dimensional Analysis}
\label{s:fda}

We propose an RCA framework that is based on the FP-Growth algorithm~\cite{Han00.1}, with multiple optimizations for production datasets in a large-scale service system. After querying the data, which is pre-aggregated using Scuba's infrastructure~\cite{Abraham13.1}, the first step in this framework is identifying the \textit{frequent item-sets} in the target state, \textit{e.g.} hardware failures or software exceptions. \textit{Item-sets} are combinations of feature values of the samples. In a structured dataset, \textit{e.g.} Table~\ref{t:structured_log_example}, the columns are considered the \textit{features} of the entities, and each feature could have multiple distinct values in the dataset. We refer to feature values as \textit{items} in the context of frequent pattern mining. For example, when analyzing hardware failures, the items could be the software configuration of the server such as the kernel and firmware versions, as well as the hardware configuration such as the device model of the various components. When analyzing software errors, the items could be the memory allocation, the machines where the jobs are run, and the version of the software package. The number of items in an item-set is called the \textit{length} of the item-set. Item-sets with greater lengths are composed of more feature values and are therefore more descriptive about the samples.

The second step in RCA is checking the strength of the associations between item-sets and the target states. We propose multiple pre- and post-processing steps for improving the scalability and the interpretability of the framework in Section~\ref{ss:performance_optimization} and \ref{ss:interpretability_optimization}.

\thispagestyle{plain} \subsection{Metrics for Evaluating the Correlations}
\label{ss:metrics}

Three main metrics are typically considered in an RCA framework: \textit{support}, \textit{confidence}, and \textit{lift}. We first describe the meaning behind these metrics in the context of root cause analysis and then describe why we picked support and lift as our main metrics to track. 

\textit{Support} was introduced by Agrawal, \textit{et al.} in~\cite{Agrawal93.1} as 
\begin{equation}\label{eq:support}
    supp(X)=\frac{|{t \in D;X \subseteq t}|}{|D|}=P(X)
\end{equation}
where $D = \{t_1, t_2, ...,t_n\}$ is a database based on a set of transactions $t_k$. 

Support of X with respect to D refers to the portion of transactions that contain X within D. In our RCA problem, $D$ is equivalent to the entire structured log, while each entry is considered a transaction $t$. Support has a \textit{downward closure} property, which is the central idea behind Apriori frequent item-set mining algorithm. Downward closure implies that all subsets of a frequent item-set are also frequent. Analogously, all supersets of an infrequent item-set can be safely pruned because they will never be frequent. The range of support is $[0, 1]$.

When mining frequent item-sets, the frequency of an item-set is defined based on the samples in the target failure state, so we limit the database to the transactions that cover the target failure state $Y$ (\textit{e.g.} software job status $=$exception). In this context, support can therefore be formulated as 

\begin{equation}\label{eq:support}
supp(X, Y)=\frac{frequency(X, Y)}{frequency(Y)}=P(X|Y)
\end{equation}
Hereafter, we refer to $supp(X)$ as the support with respect to all transactions, and $supp(X,Y)$ as the support with respect to the transactions covering $Y$.

\textit{Confidence} was introduced by Agrawal \textit{et al.} in~\cite{Agrawal93.1} and is defined as 
\begin{equation}\label{eq:confidence}
    conf(X \Rightarrow Y)=\frac{supp(X \cap Y)}{supp(X)}=P(Y|X)
\end{equation} 

Confidence, which ranges from 0 to 1, refers to the probability of $X$ belonging to transactions that also contain $Y$. Confidence is not downward closed and can be used in association rule mining after frequent item-sets are mined based on support. Confidence is used for pruning item-sets where $conf(X \Rightarrow Y) < \gamma$, where $\gamma$ is a minimum threshold on confidence. Using confidence is likely to miss good predictors for Y under imbalanced distribution of labels. For example, suppose that we have 100 failures and 1 million reference samples. If feature $X$ exists for $100\%$ of failures $Y$ but $1\%$ of references, intuitively $X$ should be a good predictor for $Y$; however confidence will be small $(<0.01)$ due to the large number of reference samples with feature $X$. For this reason we use lift in our work, which we define next.

To deal with the problems in confidence, we use the \textit{lift} metric (originally presented as \textit{interest}) introduced by Brin \textit{et al.}~\cite{Brin97.1}. Lift is defined as
\begin{equation}\label{eq:lift}
lift(X \Rightarrow Y) = \frac{conf(X \Rightarrow Y)}{supp(Y)}= \frac{P(X \cap Y)}{P(X)P(Y)}
\end{equation}

Lift measures how much more likely that $X$ and $Y$ would occur together relative to if they were independent. A lift value of 1 means independence between $X$ and $Y$ and a value greater than 1 signifies dependence. Lift allows us to address the rare item problem, whereas using confidence we may discard an important item-set due to its low frequency. A similar measure, called \textit{conviction}, was also defined in ~\cite{Brin97.1} which compares the frequency of $X$ appearing without $Y$, and in that sense it is similar to lift but conviction captures the risk of using the rule if X and Y are independent. We use lift instead of conviction primarily due to a simpler interpretation of the result for our customers.

\thispagestyle{plain} \subsection{Structured Data Logging}
\label{ss:structured_data}

\textit{Structured logs} are logs where the pieces of information in an event are dissected into a pre-defined structure. For example, in a unstructured log we may record human-readable messages about a server like the following:

\begin{lstlisting}
0:00 experienced memory error
0:00 experienced memory error
0:00 experienced memory error
0:15 reboot from tool A
0:20 experienced memory error
0:21 tool B AC Cycled the machine
0:25 no diagnosis found in tool A
0:26 tool C send to repair - undiagnosed
\end{lstlisting}

\begin{table*}[htbp]
\caption{Server errors and reboots logged in a structured table}
\label{t:structured_log_example}
\begin{center}
\begin{tabularx}{1.8\columnwidth}{cccccccc}
    \hline
    \textbf{timestamp} & \textbf{memory error} & \textbf{cpu error} & \textbf{...} & \textbf{reboot} & \textbf{undiagnosed repair} & \textbf{diagnosed repair} & \textbf{tool}\\\hline
    0:00 & 3 & 0 & & 0 & 0 & 0 & NULL \\\hline
    0:15 & 0 & 0 & & 1 & 0 & 0 & A \\\hline
    0:20 & 1 & 0 & & 0 & 0 & 0 & NULL \\\hline
    0:21 & 0 & 0 & & 1 & 0 & 0 & B \\\hline
    0:26 & 0 & 0 & & 1 & 1 & 0 & C \\\hline
\end{tabularx}
\end{center}
\end{table*}

There are a few major drawbacks in this example log. First, the same message appears multiple times, which can be aggregated and described in a more succinct way to save space. Second, tool A and tool B both write to this log, but in very different formats. Tool A and B both restarted the server by turning the power off and on, but tool A logs it as “reboot”, while tool B, developed by another group of engineers from a different background, logs it as a verb “AC Cycle”. This could even happen to the same word, for example, “no diagnosis” and “undiagnosed” in the last two messages mean the same condition, but would impose huge difficulty when one tries to parse this log and count the events with regular expressions.

With a pre-defined structure, \textit{i.e.} a list of fields to put the information in, structured logging requires a canonical way to log events. For example, in a structured table, the messages above can be logged in the format shown in Table~\ref{t:structured_log_example}.

In this conversion, engineers could decide not to log “no diagnosis found in tool A” in the structured table because it does not fit in the pre-defined structure. The structure of the table is flexible and can be tailored to the downstream application, for example, instead of having multiple columns for memory error, cpu error, \textit{etc.}, we can use one “error” column and choose a value from a pre-defined list such as memory, cpu, \textit{etc.}, to represent the same information. 

In addition to removing the ambiguity in the logs, enforcing structured logging through a single API also helps developers use and improve the existing architecture of the program, instead of adding ad-hoc functionalities for edge cases, which introduces unnecessary complexity that makes the code base much harder to maintain. In this example, if there is only one API for logging a reboot, developers from tool A and B would likely reuse or improve a common reboot service instead of rebooting the servers in their own code bases. A common reboot service would be much easier to maintain and likely have a better-designed flow to handle reboots in different scenarios.

\thispagestyle{plain} \subsection{Frequent Pattern Mining and Filtering}
\label{ss:apriori}

Our proposed RCA framework involves two steps: frequent pattern 1) \textbf{mining} and 2) \textbf{filtering}. Frequent patterns in the dataset are first reported, followed by an evaluation on how strongly each frequent pattern correlates to the target failures.

In frequent pattern mining, each item should be a binary variable representing whether a characteristic exists. In a production structured dataset, however, a column would usually represent one feature, which could have multiple distinct values, one for each entity. Therefore the structured log needs to first be transformed into a schema that fits the frequent pattern mining formulation. The transformation is done by applying \textit{one-hot encoding}~\cite{Harris12.1} on each of the columns in the structured table. For a column in the structured table $f$, which has $k$ possible values in a dataset, one-hot encoding "explodes" the schema and generate $k$ columns \{$f_0$, $f_1$, ..., $f_{k-1}$\}, each contains a binary value of whether the entity satisfies $f=f_k$.

Apriori is a classical algorithm that is designed to identify frequent item-sets. As illustrated in Algorithm~\ref{a:apriori}, starting from frequent items, \textit{i.e.} item-sets at length=1, the algorithm generates candidate item-sets by adding one item at a time, known as the \textit{candidate generation} process. At each length $k$, candidate generation is done and all the candidate item-sets are scanned to increment the count of their occurrences. Then the item-sets that meet the min-support threshold are kept and returned as the frequent item-set $L_K$. We add a practical constraint \textit{max-length} on the maximum length of the item-set that we are interested in. The limit on max-length stops the algorithm from exploring item-sets that are too descriptive and specific to the samples, which are typically less useful in production investigation.

\thispagestyle{plain} \subsection{Architecture of a Large-Scale Service Environment}

\begin{algorithm}[htbp]
\caption{Apriori Algorithm}
\label{a:apriori}
\SetAlgoLined
\DontPrintSemicolon

let $C_k$ be the candidate item-sets at length$=k$\;
let $L_k$ be the frequent item-sets at length$=k$\;
$L_1$ = frequent items\;
$k = 1$\;

\While{$L_k \neq \phi$ \textbf{and} $k \leq max\_length$}{
  $C_{k+1}$ = candidate item-sets generated from $L_k$\;
  \ForEach{transaction $t$ in database}{
    \ForEach{item-set $c$ covered by $t$}{
      increment the count of $c$\;
    }  
  }
  $L_{k+1}$ = item-sets in $C_{k+1}$ that meet min-support\;
  k++\;
}

\Return $\cup L_k$
\end{algorithm}

By generating a large set of candidates and scanning through the database many times, Apriori suffers from an exponential run time and memory complexity ($O(2^D)$), making it impractical for many production datasets. The \textit{FP-Growth} algorithm, based on a special data structure \textit{FP-Tree}, was introduced to deal with performance issues by leveraging a data structure that allows to bypass the expensive candidate generation step~\cite{Han00.1}. FP-Growth uses divide-and-conquer by mining short patterns recursively and then combining them into longer item-sets. 

Frequent item-set mining through FP-Growth is done in two phases: FP-Tree construction and item-set generation. Algorithm~\ref{a:fp_tree} shows the process of FP-Tree construction. The FP-Tree construction process takes two inputs: 1) the set of samples in the target failure state (equivalent to \textit{a transaction database} in classical frequent pattern mining literature), and 2) a min-support threshold, based on which a pattern is classified as \textit{frequent} or not. Each node in the tree consists of three fields, \textit{item-name}, \textit{count}, and \textit{node-link}. \textit{item-name} stores the item that the node represents, \textit{count} represents the number of transactions covered by the portion of the path reaching the node, and \textit{node-link} links to the next node with the same item-name. The FP-tree is constructed in two scans of the dataset. The first scan finds the frequent items and sort them, and the second scan constructs the tree.

\begin{algorithm}[htbp]
\caption{FP-Tree Construction}
\label{a:fp_tree}
\DontPrintSemicolon

Scan data and find frequent items\;

Order frequent items in decreasing order with respect to support, $F$\;

Create root node $T$, labeled as NULL\;

\ForEach{transaction $t$ in database}{
  \ForEach{frequent item $p$ in $F$}{
    \If{\text{T has a child N such that} $N$.item-set=p.item-set}{
      $|N|++$\;
    }
    \Else{
      Create $N$, link parent-link to $T$, and set $N.count = 1$\;
      Link $N$'s node-link to nodes with the same item-name\;
    }
  }
}
\end{algorithm}

Algorithm~\ref{a:fp_growth} illustrates the process for generating the frequent item-sets, based on the lemmas and properties Han \textit{et al.} proposed in \cite{Han00.1}. A \textit{conditional pattern base} is a sub-database which contains the set of frequent items co-occurring with the suffix pattern. The process is initiated by calling $FP$-$Growth(Tree, NULL)$, then recursively building the \textit{conditional} FP-Trees.

\begin{algorithm}[htbp]
\caption{Frequent Item-set Generation}
\label{a:fp_growth}
\DontPrintSemicolon
  \

  \SetKwFunction{FMain}{FB-Growth}
  \SetKwProg{Fn}{Function}{:}{}
  \Fn{\FMain{$Tree$, $\alpha$}}{

    \If{$Tree$ contains a single path $P$}{
      \ForEach{combination $\beta$ of nodes in path $P$}{
        Generate pattern $\beta \cup \alpha$ with support = min support of nodes in $\beta$\;
      }
    }
    \Else{
      \ForEach{$\alpha_{i}$ in tree}{
        Generate pattern $\beta = \alpha_{i} \cup \alpha$ with support = $\alpha_{i}$.support\;
        Construct $\beta$'s conditional pattern base and $\beta$'s conditional FP-tree $T_\beta$\;
        \If{$T_\beta \neq \phi$}{
          call \FMain{$T_\beta$, $\beta$}\;
        }
      }
    }
  }
\end{algorithm} 

After finding the frequent item-sets in the dataset, we examine how strongly the item-sets can differentiate positive (\textit{e.g.} failed hardware/jobs) samples from the negative ones. We use lift, defined in Section~\ref{ss:metrics}, to filter out item-sets that are frequent in the failure state but not particularly useful in deciding if a sample will fail. For example, an item-set can be frequent in both non-failure and failure states, and the evaluation based on lift would help us remove this item-set from the output because it is not very useful in deciding whether samples in that item-set would fail or not.

\thispagestyle{plain} \subsection{Pre- and Post-Processing for Performance Optimization}
\label{ss:performance_optimization}

We incorporated multiple optimizations as pre- and post-processing to scale the RCA framework for accommodating near real-time investigations, which are important in responding to urgent system issues quickly. Many entities in a production log are identical, except the columns that are unique identifiers of the entities such as the timestamps, hostnames, or job IDs. Utilizing Scuba's scalable infrastructure~\cite{Abraham13.1}, we query \textit{pre-aggregated} data which are already grouped by the distinct combinations of column values, with an additional \textit{weight} column that records the count of the identical entities. To handle this compact representation of the dataset, we modified the algorithms to account for the weights. This pre-aggregation significantly reduces the amount of data that we need to process in memory and would reduce the runtime of our production analyses by $> 100X$.

Columns that are unique identifiers about the entities need to be excluded before the Scuba query. The aggregation in Scuba is only meaningful after excluding these columns, otherwise the aggregation would return one entity per row due to the distinct values per entity. The framework allows users to specify columns to be excluded in the dataset, as well as automatically checks to exclude columns with the number of distinct values $>D$ portion of the number of samples. Empirically, we use $D=2\%$ in one of our applications, and the proper setting of $D$ highly depends on the nature of the dataset.

Adding multithreading support to the algorithm further improves Apriori's performance, as the algorithm generates a large number of combinations and test them against the data. By testing these combinations in parallel, we can scale up with the number of available cores. However, we found that FP-Growth outperforms Apriori even when Apriori is optimized with multithreading.

\thispagestyle{plain} \subsection{Interpretability Optimization}
\label{ss:interpretability_optimization}

In production datasets, it is common that there exist a large number of distinct items, and the lengths of the findings are typically much smaller than the number of the one-hot encoded feature columns (as discussed in Section~\ref{ss:apriori}). As a result, there can be multiple findings describing the same group of samples. To improve the quality of the result, we implemented two filtering criteria for removing uninteresting results as described below:

\vspace{5mm}
\textbf{Filter 1}: An item-set $T$ is dropped if there exists a proper subset $U, (U \subset T)$ such that $lift(U)*H_{lift} \geq lift(T)$, where $H_{lift} \geq 1$ 
\vspace{5mm}

If there exist shorter rules with similar or higher lift, longer rules are pruned because they are less interesting. $H_{lift}$ is a multiplier that can be tuned based on the nature of the dataset, to remove more longer rules, as it makes the condition easier to be satisfied. This filter addresses the ubiquitous items discussed in~\cite{Bittmann18.1}. As there exists a shorter rule with similar or higher lift, the one containing the ubiquitous item will be filtered out. Consider two rules: 

\begin{lstlisting}
{kernel A, server type B} => failure Y with lift 5 
{kernel A, server type B, datacenter C} => failure Y with lift 1.5
\end{lstlisting}

It is likely that describing the server and kernel interaction is more significant than filtering by datacenter, therefore the second rule is pruned, even though the lift values from both rules meet our threshold on the minimum lift.

\vspace{5mm}
\textbf{Filter 2}: An item-set $T$ is dropped if there exists a proper superset $S, (S \supset T)$ such that $supp(S)*H_{supp} \geq supp(T)$ and $lift(S) > lift(T)*H_{lift}$, where $H_{supp} \geq 1$ and $H_{lift} \geq 1$ 
\vspace{5mm}

If a rule has a superset which describes a similar number of samples, \textit{i.e.} similar support, and the superset's lift is higher, the rule will be dropped as the superset is a more powerful rule that describes most or all of the samples. Similarly, $H_{supp}$ is applied to loosen the comparison criteria for support, and $H_{lift}$ is applied to ensure that the lift difference is sufficiently large based on the use case. For example, consider two rules:

\begin{lstlisting}
{datacenter A} => failure Y with support 0.8 and lift 2
{datacenter A, cluster B, rack C} => failure Y with support 0.78 and lift 6
\end{lstlisting}

In this scenario, almost all of the samples in datacenter A are also in cluster B and rack C. When trying to understand the root cause of the failures, the longer item-set with a higher lift and a similar support is more informative, so we keep it and remove the shorter item-set.

In summary, we prefer to keep shorter item-sets when the lift values are similar. When a longer item-set's lift value is significantly higher than that of a subset, and the support values are similar, we keep the longer item-set.

\thispagestyle{plain} \subsection{The Proposed Framework}
\label{ss:proposed_framework}

\begin{figure}[htbp]
  \centering
  \includegraphics[width=5.5cm]{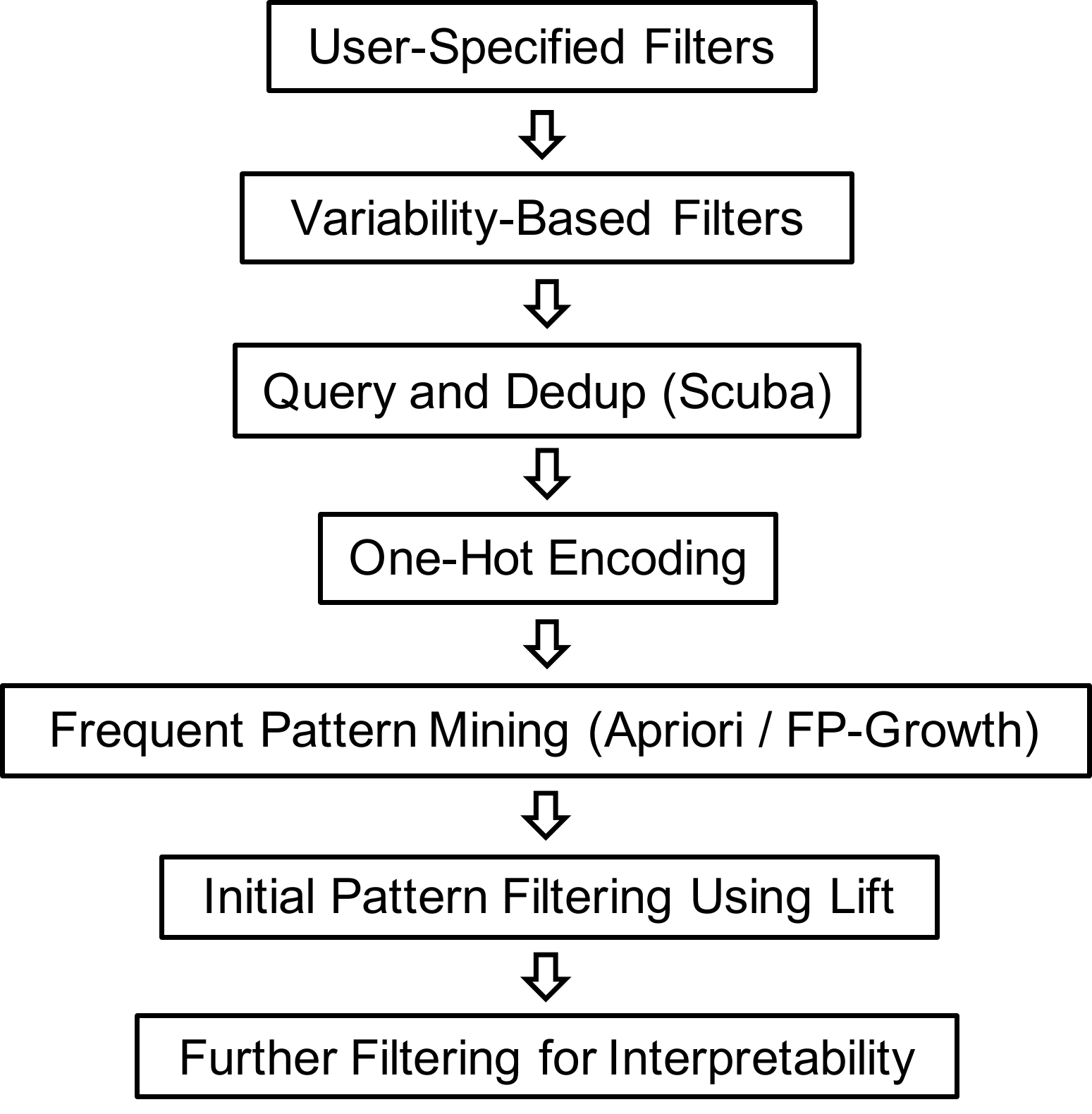}
  \caption{The proposed framework for the Fast Dimensional Analysis.}
  \label{f:flow}
\end{figure}

Our final implementation incorporating the above-mentioned improvements is illustrated in Figure~\ref{f:flow}. Utilizing Scuba's scalable, in-memory infrastructure~\cite{Abraham13.1}, we query data that is aggregated by the \textit{group-by} operation based on the distinct value combinations in a set of columns. The aggregation is done excluding columns that the users specify as not useful, and columns that our check finds to have too many distinct values to satisfy a practical minimum support threshold. One-hot encoding is then applied to the queried data for converting the column-value pairs to Boolean columns, or items. We apply frequent pattern mining techniques such as Apriori and FP-Growth on the dataset to identify frequent item-sets, which are then filtered by lift because in RCA we are only interested in item-sets that are useful in separating the target labels, \textit{e.g.} specific failure states, from the rest of the label values, \textit{e.g.} successful software task states. Finally the filtering criteria in Section~\ref{ss:interpretability_optimization} are applied to further condense the report for better interpretability.

In our framework we choose to filter the association rules by lift and support after the rules are mined with FP-Growth, instead of pruning the association rules during the tree construction as the STUCCO algorithm does in a contrast set mining problem~\cite{Bay99.1,Bay01.1,Castelluccio17.1}. This helps us find more granular item-sets with high lift that would otherwise be missed. For example, if association rule $\{A, B\} \Rightarrow Y$ has a high lift (or the $\chi^2$ statistic as in STUCCO), but both $\{A\} \Rightarrow Y$ and $\{B\} \Rightarrow Y$ have lift (or $\chi^2$ statistic) values below the pruning threshold, $\{A, B\} \Rightarrow Y$ would not be found if we prune the tree based on both support and lift (or $\chi^2$ statistic) as STUCCO does. In STUCCO every node needs to be \textit{significant} based on chi-square tests and \textit{large} based on support for it to have child nodes~\cite{Bay01.1}. On the other hand, as FP-Growth mines frequent item-sets only based on support, as long as $\{A\} \Rightarrow Y$, $\{B\} \Rightarrow Y$, and $\{A, B\} \Rightarrow Y$ have enought support, they would all be reported as frequent item-sets. Then $\{A\} \Rightarrow Y$ and $\{B\} \Rightarrow Y$ will be filtered out due to low lift while $\{A, B\} \Rightarrow Y$ will be reported in the final result.

\section{Experimental Results}
\label{s:experimental_result}

In this section we present the optimization result for runtime and interpretability, based on their relationships with the min-support and max-length parameters during item-set mining, as well as the min-lift during item-set filtering. We experimented on two production datasets:

\begin{itemize}
\item \textit{Anomalous Server Events}

This is a dataset about anomalous behaviors (\textit{e.g.} rebooted servers not coming back online) on some hardware and software configurations (more details can be found in Section~\ref{ss:anomalous_hw_and_sw}). We compiled the dataset with millions of events, each described by 20 features. The features expand to about 1500 distinct items after one-hot encoding, with tens of thousands of distinct valid item-sets. Approximately $10\%$ of the data are positive samples, \textit{i.e.} anomalous behaviors. For simplicity we refer to this dataset as \textit{ASE} in the rest of the paper. 

\item \textit{Service Requests} 

This is a dataset that logs the requests between services in a large-scale system. Each request is logged with information such as the source and destination, allocated resource, service specifications, and authentication (more details can be found in Section~\ref{ss:anomalous_service_interactions}). For experimentation, we compiled a dataset that contains millions of requests, each described by 50 features. The features expand to about 500 distinct items after one-hot encoding, with 7000 distinct valid item-sets. Approximately $0.5\%$ of the data are positive samples. For simplicity we refer to this dataset as \textit{SR} in the rest of the paper. 
\end{itemize}

As the datasets are different by an order of magnitude in terms of the proportion of positive samples, we expect our range of lift to vary considerably. Additionally, the ASE dataset contains almost $5X$ the number of distinct feature values as the SR dataset does, which would affect the count and length of item-sets mined with respect to min-support. We demonstrate the results based on these two datasets with the different characteristics below.

\thispagestyle{plain} \subsection{Performance Improvement}
\label{ss:performance_improvement}

\subsubsection{Data Pre-Aggregation}
\label{sss:data_pre-agg}

As described in Section~\ref{ss:performance_optimization}, many events in production logs are identical, except for the unique identifiers such as job IDs, timestamps, and hostnames. The pre-aggregation of the data could effectively reduce the size of the ASE dataset by $200X$ and the SR dataset by $500X$, which significantly improves both the data preparation time and the execution time of the FP-Growth algorithm. 

Table~\ref{t:data_prep_time} shows the runtime needed for querying the data and grouping the entries with and without the pre-aggregation using Scuba's infrastructure. Without the pre-aggregation step in Scuba, the data is queried from Scuba and grouped in memory before our framework consumes it. Without pre-aggregation, the large amount of data that is transferred has a significant performance impact on the query time. Overall the data preparation time is improved by $10X$ and $18X$ for the ASE and SR datasets.

\begin{table*}[bhtp]
\caption{Data preparation time with and without pre-aggregation in Scuba (in seconds)}
\label{t:data_prep_time}
\begin{center}
\begin{tabularx}{1.9\columnwidth}{cccc}
    \hline
    & \textbf{Scuba query time (including pre-agg. time)} & \textbf{Grouping time (in memory)} & \textbf{Total time} \\\hline
    ASE (without pre-agg.) & 3.00 & 0.94 & 3.94 \\\hline
    ASE (with pre-agg.) & 0.39 & - & 0.39 \\\hline
    SR (without pre-agg.) & 3.35 & 0.71 & 4.06 \\\hline
    SR (with pre-agg.) & 0.22 & - & 0.22 \\\hline
\end{tabularx}
\end{center}
\end{table*}

After the data is grouped, we execute our modified FP-Growth algorithm which takes the count of samples per unique item-set, as a weight for additional input. The details of the modification is discussed in Section~\ref{ss:performance_optimization}. The additional weight value is used in calculating the support and lift of an item-set, and has negligible overhead on the algorithm's runtime. Effectively, this means the algorithm now only need to handle $200X$ and $500X$ fewer samples from the ASE and SR datasets. Hence, the pre-aggregation of the data is critical for enabling automatic RCA at this scale in \textit{near real-time}. This is one of the major features of this paper, as most of the prior works mentioned in Section~\ref{ss:prior_work} did not incorporate any optimization using a production data infrastructure, and could not handle large-scale datasets in near real-time.

\subsubsection{Optimizing for Number of Item-Sets}

\begin{figure}[htbp]
\begin{center}
    \begin{subfigure}{\columnwidth}
      \includegraphics[width=8cm]{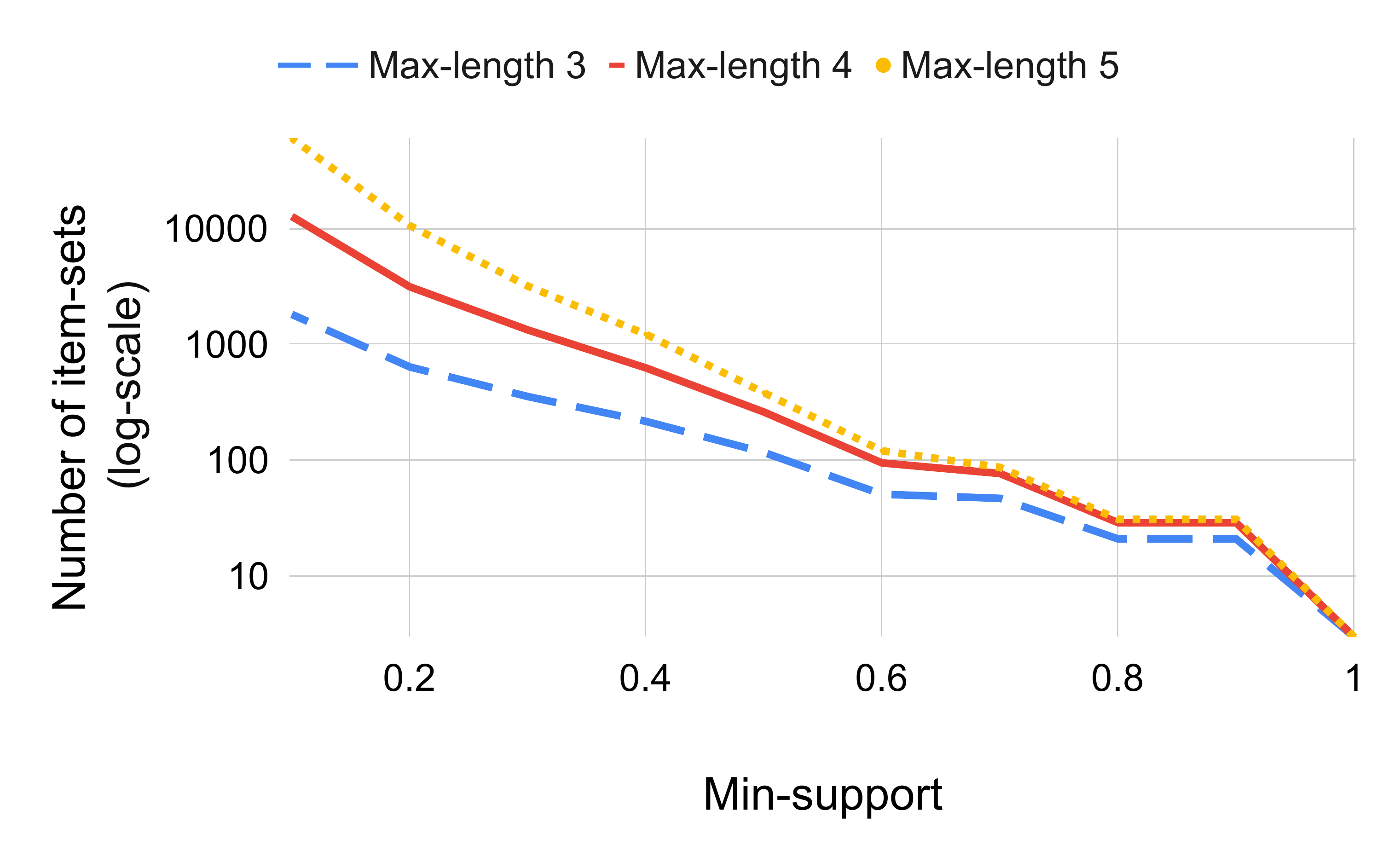}
      \subcaption{ASE dataset}
      \label{f:itemset_vs_min_support_reboot}
    \end{subfigure}\\%
    \vspace{30pt}
    \begin{subfigure}{\columnwidth}
      \includegraphics[width=8cm]{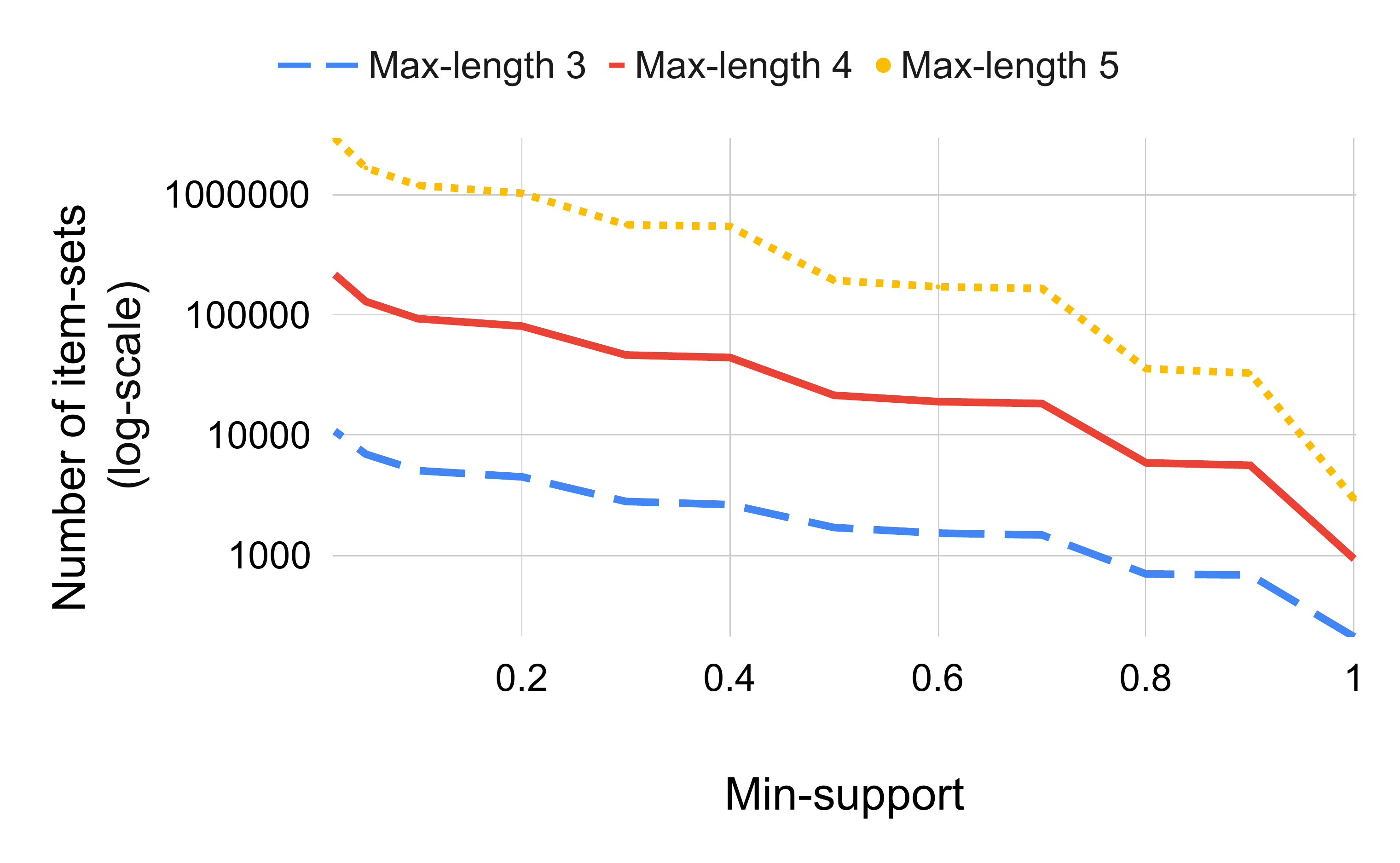}
      \subcaption{SR dataset}
      \label{f:itemset_vs_min_support_sr}
    \end{subfigure}
    \caption{Relationship between number of mined item-sets and min-support and max-length.}
    \label{f:itemset_vs_min_support}
\end{center}
\end{figure}

Item-set generation is the biggest factor in runtime of the proposed framework. We first examine the relationship between the number of reported frequent item-sets and min-support and max-length in Figure~\ref{f:itemset_vs_min_support}. Note that the vertical axes are in log scale.

\thispagestyle{plain}
\vspace{20pt}

In Figure~\ref{f:itemset_vs_min_support_reboot}, based on the anomalous server event (ASE) dataset, we see an exponential decrease in the number of item-sets as we increase min-support. Theoretically, the number of candidate item-sets is bounded by $\sum_{k=1}^{max-length} {\text{number of items} \choose k}$. In practice, however, the number of candidates is much lower because many items are mutually exclusive, \textit{i.e.} some item-sets would never exist in production. The number of item-sets based on the three max-length values converge to within an order of magnitude when min-support is around $0.4$, meaning a greater proportion of item-sets with support greater than $0.4$ can be covered by item-sets at length $3$.

Checking the convergence point helps us decide the proper max-length, given a desired min-support. For example, in a use case where the goal is to immediately root cause a major issue in production, we would be interested in item-sets with higher supports. In the ASE example, if our desired min-support is greater than the convergence point in Figure~\ref{f:itemset_vs_min_support_reboot}, say $0.95$, we only need to run the analysis with max-length set to 3, to avoid unnecessary computations for optimized performance. 

On the other hand, if the goal is to thoroughly explore the root causes for smaller groups, with less concerns about the runtime, we could set the min-support to a smaller value such as $0.4$. In this case, max-length should be set sufficiently high so that second filter discussed in Section~\ref{ss:interpretability_optimization} can be effective to improve interpretability. For example, if a rule of interest is described by $\{A, B, C, D, E\} \Rightarrow Y$, but max-length is smaller than $5$, up to ${5 \choose max-length}$ item-sets could be generated to represent this rule at the same support, whereas if max-length is set to $5$, the exact rule of interest would be created and the rest item-sets with smaller lengths would be dropped by the second filter in Section~\ref{ss:interpretability_optimization}, as long as the rule of interest has a higher lift.

The same trends are observed in Figure~\ref{f:itemset_vs_min_support_sr} for the service request (SR) dataset when increasing min-support or max-length, but there does not exist a clear convergence point. Additionally, the number of item-sets are non-zero when min-support is as high as $1$, implying there are multiple item-sets with support being $1$ at the different max-lengths. In practice, these item-sets with support being $1$ often could be represented by more specific item-sets, \textit{i.e.} supersets, and therefore could be filtered out by the filters in Section~\ref{ss:interpretability_optimization} if the supersets have higher lift. Figure~\ref{f:itemset_vs_min_support_reboot} and Figure~\ref{f:itemset_vs_min_support_sr} demonstrate that the relationship between the number of mined item-sets and min-support is dataset-dependent, and the convergence point of different max-lengths determined by the complexity of the datasets.

\subsubsection{Runtime Improvement}

\begin{figure}[htbp]
  \centering
  \includegraphics[width=8cm]{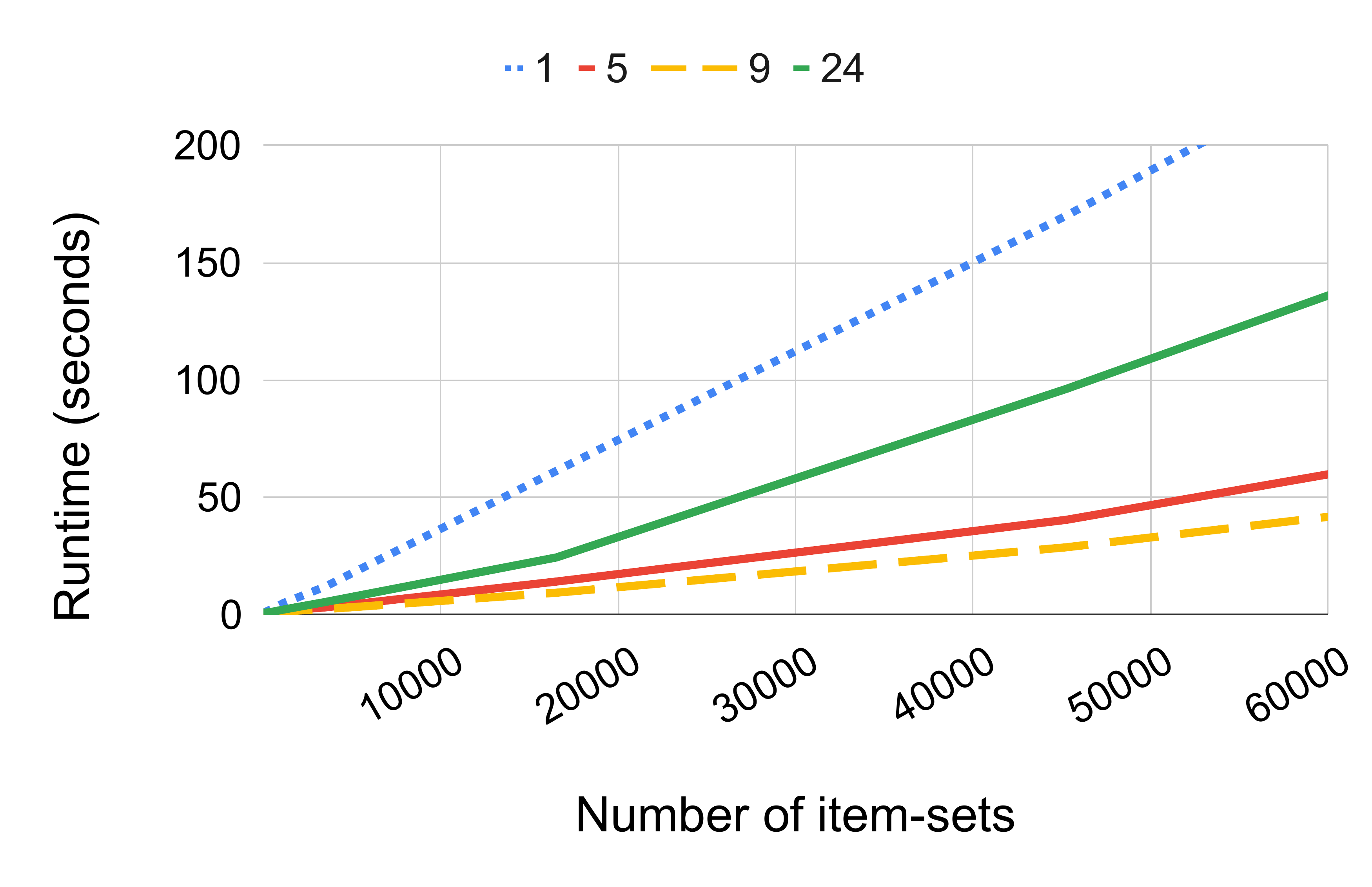}
  \caption{Runtime of Apriori at different thread counts, based on the ASE dataset.}
  \label{f:threads}
\end{figure}

An advantage of Apriori is that it is easily parallelizable, by splitting up the candidate generation at each length. The optimal parallelism level depends on the number of candidates, since each thread induces additional overhead. Figure~\ref{f:threads} illustrates the runtime of Apriori at different levels of parallelism, based on the ASE dataset. The runtime is reported based on a machine with approximately 50 GB memory and 25 processors. As shown in the figure, a 9-thread parallelism resulted in the shortest runtime due to a better balance between the runtime reduction from parallel processing and the runtime increase from parallelization overheads. Every parallelism level up to 24 threads outperforms the single-threaded execution. 

\thispagestyle{plain}

\begin{figure}[htbp]
\begin{center}
    \begin{subfigure}{\columnwidth}
      \includegraphics[width=8cm]{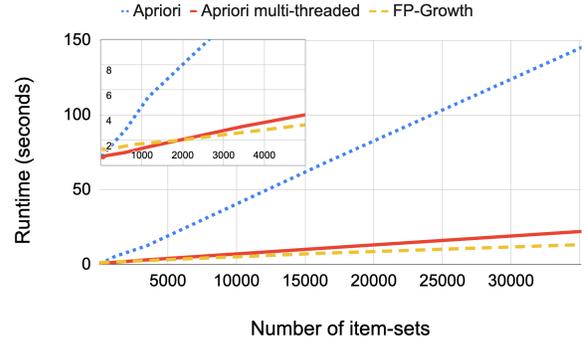}
      \subcaption{ASE dataset}
      \label{f:runtime_vs_itemset_combined_reboot}
    \end{subfigure}\\%
    \vspace{30pt}
    \begin{subfigure}{\columnwidth}
      \includegraphics[width=8cm]{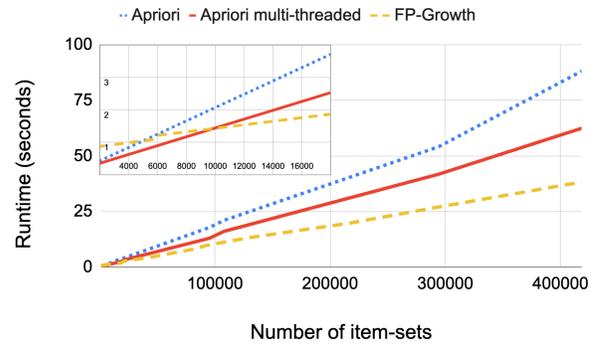}
      \subcaption{SR dataset}
      \label{f:runtime_vs_itemset_combined_sr}
    \end{subfigure}
    \caption{Analysis runtime vs. number of item-sets.}
    \label{f:runtime_vs_itemset_combined}
\end{center}
\end{figure}

In Figure~\ref{f:runtime_vs_itemset_combined}, we compare the runtime of Apriori~\cite{Agrawal93.1}, parallelized Apriori, and FP-Growth~\cite{Han00.1,Arifin16.1,Liu16.1,Wang17.1,Bittmann18.1}, at different number of item-sets. Note that the 9-thread configuration from Figure~\ref{f:threads} is used as the multi-threaded case here. It is clear that FP-Growth outperforms single-threaded and multi-threaded Apriori, except when the number of item-sets is small, as the overhead of setting up the FP-Growth algorithm (see Section~\ref{ss:apriori}) is larger than the benefit of not running Apriori's candidate generation step. However, in our experiments, FP-Growth is never slower for more than 1 second. The runtime difference could become more significant when this RCA framework is deployed on a resource-limited platform, such as an embedded system. In Figure~\ref{f:runtime_vs_itemset_combined_sr}, we see that for the SR dataset, Apriori can be faster when the number of item-sets is smaller than $10000$, which happens when max-length $<4$ or (max-length $=4$ and min-support $\geq 0.8$). For the ASE dataset, multi-threaded Apriori is faster than FP-Growth when the number of item-sets is smaller than 2000, which happens when min-support $\geq 0.4$. For a given dataset, running an initial scan over the algorithms, lengths, and supports can help optimize the choice of algorithm and min-support. 

\thispagestyle{plain} \subsection{Interpretability Improvement}
\label{ss:interpretability_improvement}

\begin{figure}[htbp]
\begin{center}
    \begin{subfigure}{\columnwidth}
      \includegraphics[width=8cm]{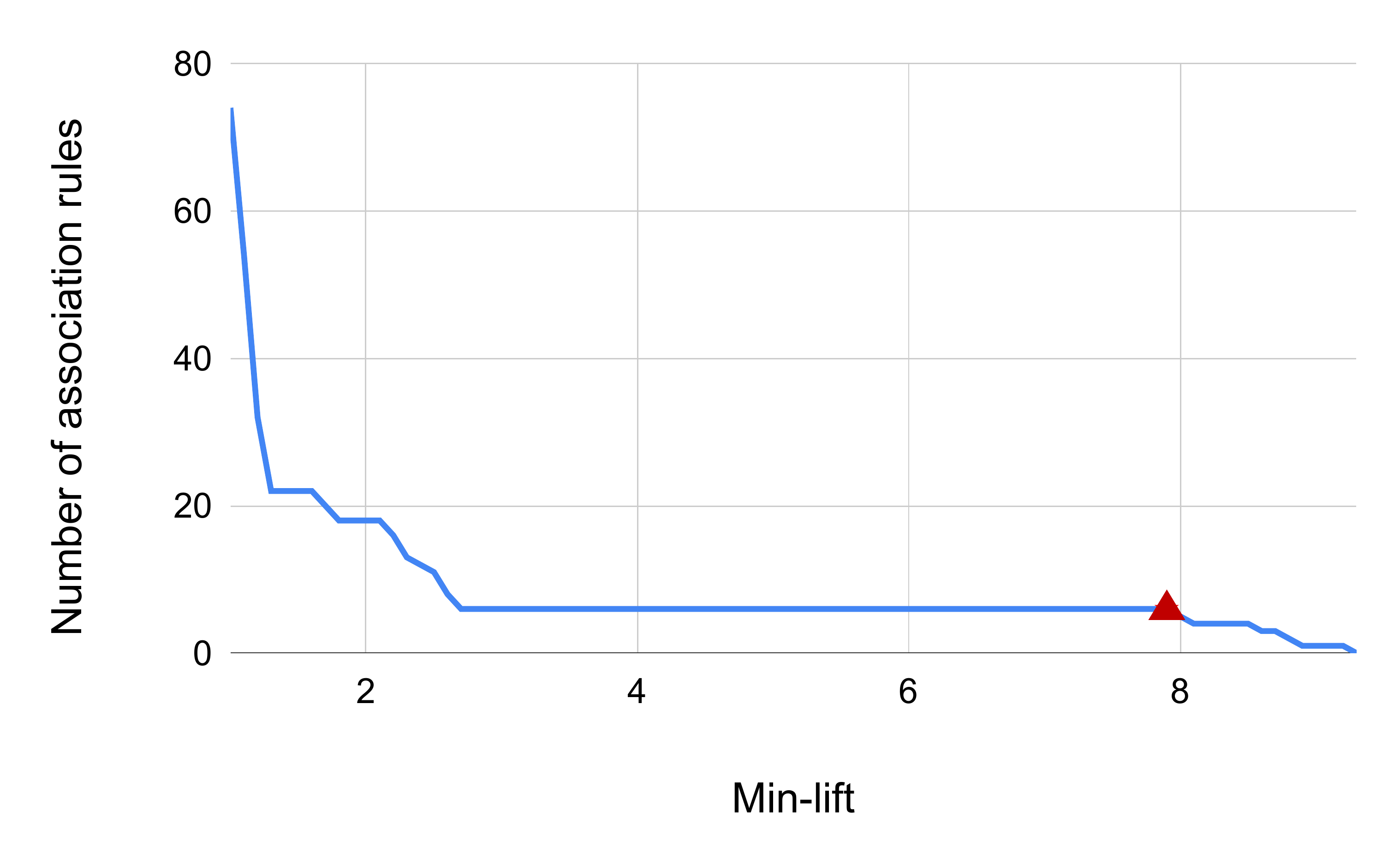}
      \subcaption{ASE dataset}
      \label{f:assoc_rule_vs_min_lift_reboot}
    \end{subfigure}\\%
    \vspace{30pt}
    \begin{subfigure}{\columnwidth}
      \includegraphics[width=8cm]{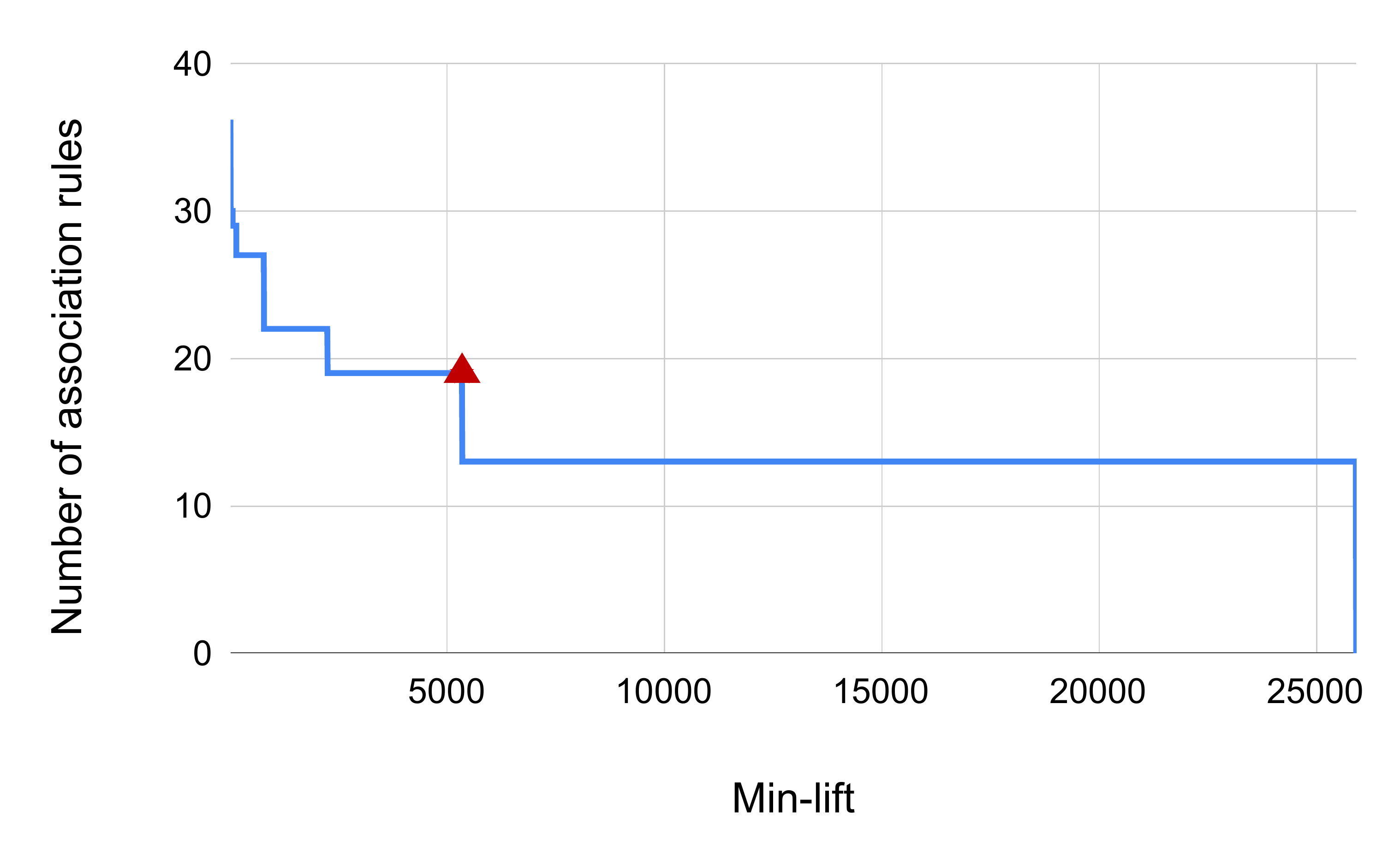}
      \subcaption{SR dataset}
      \label{f:assoc_rule_vs_min_lift_sr}
    \end{subfigure}
    \caption{Number of reported association rules vs. min-lift threshold. The triangle marks when a target rule appears in the use cases discussed in Section~\ref{s:use_case}.}
    \label{f:assoc_rule_vs_min_lift}
\end{center}
\end{figure}

As described in Section~\ref{ss:metrics}, lift is used when filtering the mined frequent item-sets based on their ability in deciding whether a sample satisfying the item-set would be positive (\textit{e.g.} be in the target failure states). Figure~\ref{f:assoc_rule_vs_min_lift_reboot} shows the number of association rules given different min-lift thresholds on the ASE dataset, when min-support is set to $0.4$ and max-length is set to $5$.

There is a clear drop after min-lift $=1$, which indicates rules that are stronger than randomness. The number of association rules remains constantly at $6$ when min-lift is between $2.7$ and $7.9$. In practice, we can set the min-lift to anywhere between $2.7$ and $7.9$ to output these $6$ rules as the potential root causes, as they are the relatively stronger and more stable rules in the dataset. The triangle marker indicates when an example actionable insight appears at the highest min-lift value (more discussions in Section~\ref{ss:anomalous_hw_and_sw}).

Figure~\ref{f:assoc_rule_vs_min_lift_sr} shows the same analysis based on the SR dataset, with min-support set to $0.5$ and max-length set to $5$. The number of association rules reported drops significantly in several steps. This is because there does not exist a clear convergence point for different max-length values, as seen in Figure~\ref{f:itemset_vs_min_support_sr}, many of the reported association rules actually describe the same underlying rules, and therefore are filtered out together as min-lift increases. The triangle marker shows when an example actionable insight appears at the highest min-lift value (more discussions in Section~\ref{ss:anomalous_service_interactions}). Compared to the ASE dataset, the lift is much larger in the SR dataset.

\thispagestyle{plain}
\vspace{20pt}

\begin{figure}[htbp]
  \centering
  \includegraphics[width=8cm]{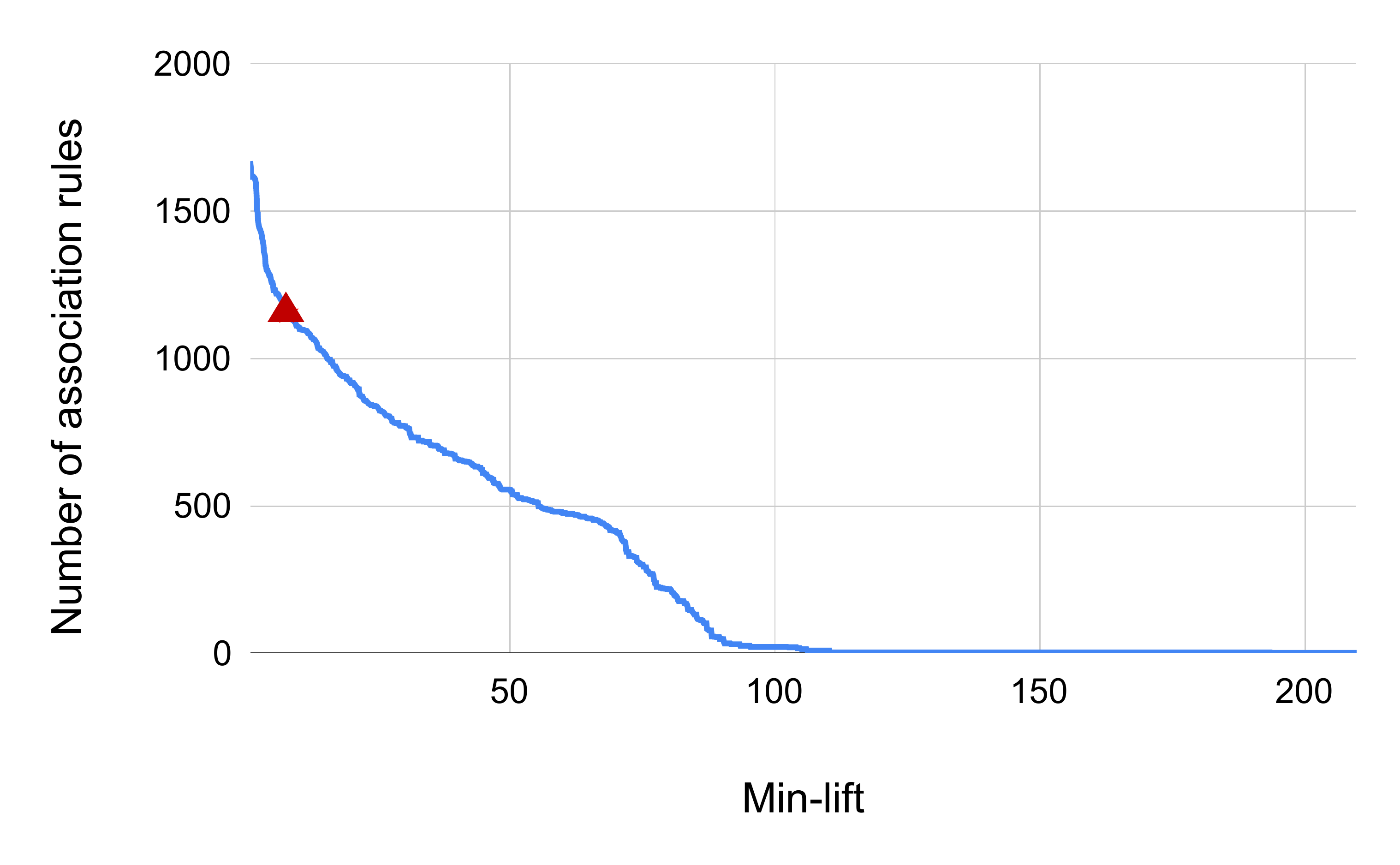}
  \caption{Number of association rules at different min-lift thresholds based on the ASE dataset, when min-support is set to 0.1. The triangle marks when a target rule appears in the use case discussed in Section~\ref{s:use_case}.}
  \label{f:assoc_rule_vs_min_lift_reboot_low_support}
\end{figure}

To understand the larger trend across all association rules in the ASE dataset, we consider more item-sets by lowering min-support to $0.1$. An exponentially decreasing trend can be observed in Figure~\ref{f:assoc_rule_vs_min_lift_reboot_low_support}. For reference, we kept the same triangle marker at min-lift $= 7.9$, representing a highly actionable insight confirmed by service engineers. This graph also illustrates the importance of setting a sufficiently high min-support to reduce noise. When using min-support 0.4 derived from Figure~\ref{f:assoc_rule_vs_min_lift_reboot}, we have six rules above lift 7 compared to 1200 for min-support 0.1.

\thispagestyle{plain}

With the two post-processing filters described in Section~\ref{ss:interpretability_optimization}, at max-length $=5$ and min-lift $=1$, we kept only $2\%$ and $26\%$ of the mined rules from the ASE dataset when min-support is $0.1$ and $0.4$, while we kept only $0.4\%$ and $1\%$ of the mined rules from the SR dataset when min-support is $0.1$ and $0.5$. The post-processing reduces the number of mined association rules significantly without losing the important root cause information, which makes the output report much easier for human to examine and debug quickly. While the work in~\cite{Liu16.1} integrates the FP-Growth algorithm with the Spark platform for processing large datasets and uses lift for pruning mined item-sets, it did not provide well-formulated post-processing steps or characterize the relationship between min-lift and the mined item-sets. 

\subsection{Lessons Learned}
\label{ss:lessons}

While the results presented in this paper demonstrate effective RCA, in practice, the relationships between these parameters are highly dependent on the nature of the datasets. Therefore we present methods for optimizing the performance and interpretability. First of all, min-support is a variable that controls how granular the reported rules would be. In a real-time analysis or debug, a lower max-length can be set to reduce runtime, and a higher min-support can be applied to report only the most dominant issues in the dataset. In a more thorough analysis that is less time-sensitive, a higher max-length can be applied to generate a set of rules that are overall more descriptive, based on the filtering criteria in Section~\ref{ss:interpretability_optimization}.

If there exists a clear convergence point given different max-lengths, a lower max-length should be used to avoid unnecessary computation. If the RCA application is very sensitive to runtime and the number of item-sets is small, one could first run the analysis similar to the one presented in Figure~\ref{f:runtime_vs_itemset_combined} and use multi-threaded Apriori in the region where it outperforms FP-Growth.

The advantage of support and lift is that they are very interpretable and intuitive metrics that any service engineer can adjust. One intuition behind the lift value is to make sure we handle the edge case where a label value, \textit{e.g.} specific failure states, has attribution $X$, and no other label values has attribution $X$.

\section{Use Case Study}
\label{s:use_case}

The method discussed in this paper has been productionized in multiple hardware, software, and tooling applications in our large-scale service infrastructure. Deployment of this framework allows for fast root cause analysis as well as automatic alerting on new correlations in the datasets, which may indicate unexpected changes in the systems. In this section we present some of the applications and the insights (after sanitizing the data) that were extracted by the proposed framework.

\thispagestyle{plain} \subsection{Anomalous Hardware and Software Configurations}
\label{ss:anomalous_hw_and_sw}

In a large infrastructure, maintenance activities are constantly undertaken by the management system - for instance, we might need to provision new services on a particular server platform. In such scenarios, there might be reasons to reboot servers. One root cause example here is to detect whether all servers have booted back up after a maintenance event. Using our framework, we found a group of servers that failed to come back online as compared to the rest of the cohorts. Without our proposed root cause analysis, the issue was isolated to a combination of 1) a specific firmware version in one component, 2) a particular component model from a manufacturer, and 3) a particular server model, by experienced experts after hours of investigation.

To emulate how the proposed fast dimensional analysis could have helped with the root cause analysis, we looked at the historical data and labeled the servers by whether the reboots were successful on them. For example, since the servers that stayed offline is our target of the investigation, we labeled them as positive, and the rest where the reboots were successful as negative. Then we compiled a dataset that joins the labels with about 20 attributes of the servers, such as the server model, the type of services the servers were running, firmware and kernel versions, component vendors/models/firmware versions. These attributes are where we expect to find potential correlations for distinguishing between the positive and negative samples. This is the first dataset presented in the experimental results in Section~\ref{s:experimental_result}, \textit{i.e.} the anomalous server event (ASE) dataset.

With this dataset, the fast dimensional analysis framework identified the correlation based on exactly the three attributes in 2 seconds. The lift value where this target association rule shows up is marked by the triangle in Figure~\ref{f:assoc_rule_vs_min_lift_reboot}. Through our methodology, we significantly reduced the investigation time from hours to seconds. Note that in this case, there were multiple combinations of feature values that correlate to the positive samples equally. For example, a combination of \textit{\{firmware version, component model, server model\}} would show the same support and lift as a combination of \textit{\{storage interface, component model, CPU model\}}, on this specific dataset. Purely based on this dataset, the algorithm would not be able to tell which combination is more useful given the type of failures. Further analysis can determine the most effective way to reduce the number of combinations reported, potentially based on the past reports. The reported combinations already provides strong correlations to the failures and an engineer with some experience can quickly conclude the issue from the report.

\thispagestyle{plain} \subsection{Anomalous Service Interactions}
\label{ss:anomalous_service_interactions}

All the communications between backend services in our large-scale system are logged. This information is used to investigate errors in the communication among services, based on characteristics such as latency, timeouts, requests, responses, traffic (volume, source and destination regions). This is the second dataset, \textit{i.e.} the service request (SR) dataset presented in the experimental results in Section~\ref{s:experimental_result}.

The naive investigation where engineers aggregate the various parameters through a group-by operation does not scale, as there are too many distinct combinations of the column values. We deployed the fast dimensional analysis framework to analyze two types of anomalous service interactions: errors and latency. The analysis quickly identified attributes of service communication that would lead to different types of errors and reported the findings. In one example for a globally distributed service, it was reported that the errors were caused only for communications between two specific geographical locations. This prompted engineers to investigate in this direction and fix the issue timely. An actionable insight based on \textit{\{service type, build version\}} $\Rightarrow$ \textit{failure} is marked by the triangle in Figure~\ref{f:assoc_rule_vs_min_lift}.

Latency is not discrete when compared to errors, hence we need to first bucketize latency values into a finite number of intervals, \textit{e.g.} acceptable and non-acceptable latencies. The framework then identifies the combinations of features where requests have non-acceptable latencies. By tuning the bucketing threshold we obtained insightful correlations based on the features of the service requests, which are used to optimize the performance of the systems.

\thispagestyle{plain} \subsection{Failed Auto-Remediations}
\label{ss:failed_auto_remediation}

We deployed the fast dimensional analysis framework on the logs from an auto-remediation system~\cite{Isard07.1,Verma15.1,Lin18.1} to quickly identify attributes of the remediation jobs that would lead to different types of exceptions, and report the correlations to a visualization dashboard that engineers use everyday for monitoring system health. For analyzing the correlations in auto-remediation jobs, we prepare about 20 server attributes mentioned above, and join them with some basic information of the remediation jobs such as failure mode, remediation name, and owner of the remediation. 

Different from the previous example, where server reboots would either succeed or fail, the auto-remediation jobs could end up in different states. In addition to successfully fixing the issue, remediation jobs could end up as \textit{repair}, \textit{i.e.} the hardware failure needs a physical repair, \textit{escalate}, \textit{i.e.} the failure needs to be escalated to human even before the system creates a repair ticket for it, \textit{rate limited and retry}, \textit{i.e.} the remediation is temporarily suspended because there are too many servers going through remediation at the time, and \textit{exception}, \textit{i.e.} the job encounters some exception and could not finish.

As the auto-remediation system serves hundreds of different remediations for a complex set of failure types, there are typically failed remediations, \textit{i.e.} escalates and exceptions, in production. The problem formulation here is hence different. Instead of finding correlations to a single issue, as we did in root causing the failed reboots in the previous example, here we explore strong correlations among the many types of failures that are constantly happening. Since the auto-remediation system is centralized and processes the remediation of the entire whole of machines, a small portion of the overall remediations may mean the majority for a small service, and the service may actually have large impact to the entire service infrastructure. Therefore, in this setup we chose a much smaller threshold on support, and report all high-lift correlations to service owners for investigation. With this setup, we have been able to identify strong correlations such as \textit{\{kernel version, service\}} $\Rightarrow$ \textit{exception} and \textit{\{service type, remediation name\}} $\Rightarrow$ \textit{escalate}, which helped engineers quickly identify and fix problems in the systems. 

\begin{figure}[htbp]
  \centering
  \includegraphics[width=8cm]{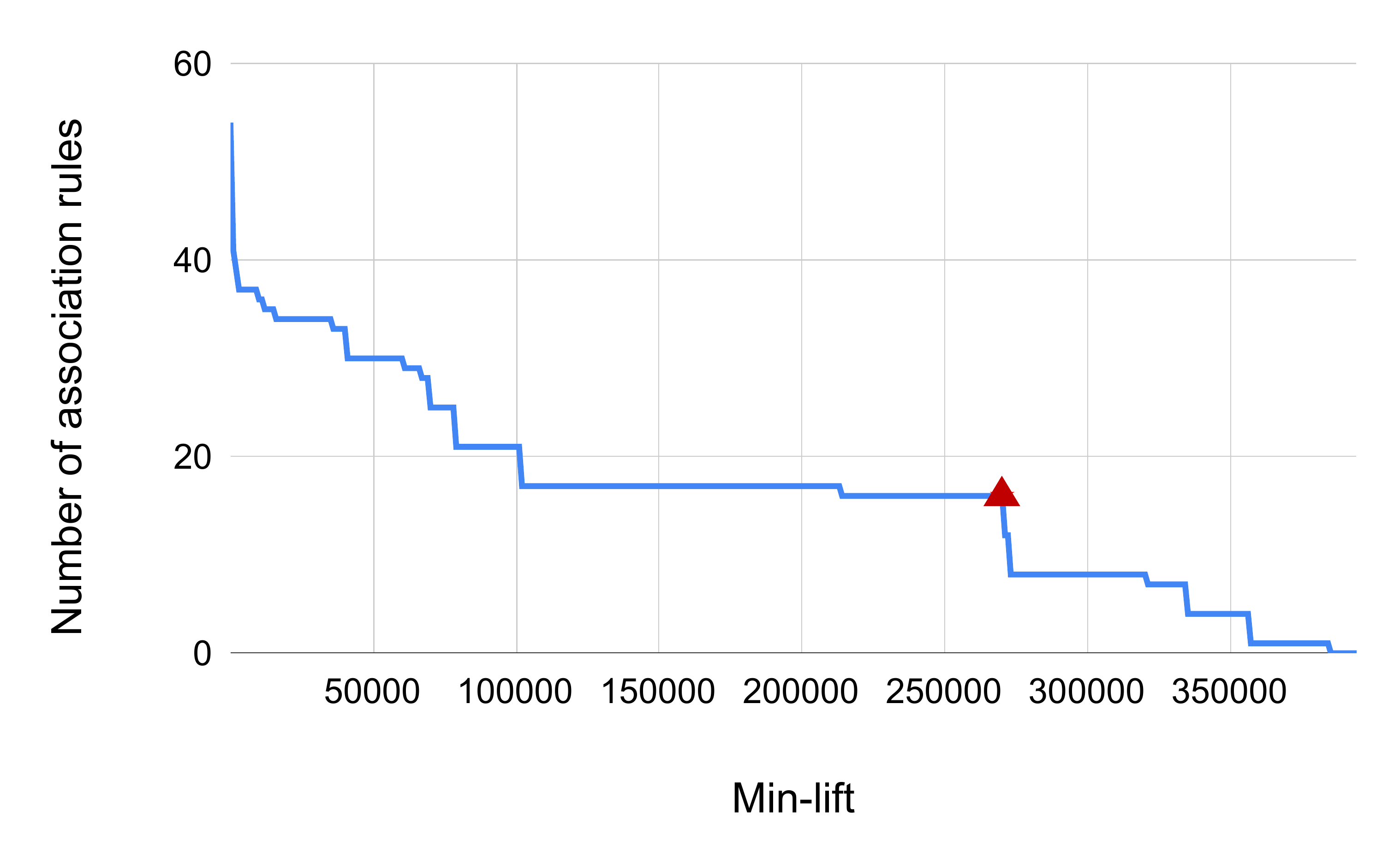}
  \caption{Number of association rules at different min-lift thresholds based on the auto-remediation dataset. The triangle marks when a target rule appears in the use case.}
  \label{f:assoc_rule_vs_min_lift_fbar}
\end{figure}

The number of association rules reported at different min-lift values is plotted in Figure~\ref{f:assoc_rule_vs_min_lift_fbar}, where a target rule mentioned above, \textit{\{service type, remediation name\}} $\Rightarrow$ \textit{escalate}, is found when min-lift is $\leq 270000$, marked by the triangle.

\thispagestyle{plain} \subsection{SSH Logs}
\label{ss:ssh_logs}

We deployed the fast dimensional analysis framework on a dataset containing logs for SSH connections to identify what causes some of the connections to fail. We passed a number of attributes from this dataset, such as the source server type, destination server type, and SSH method, in order to find out root cause connection failures. The report is exported to a dashboard for the engineers to continuously monitor the correlated rules, and to quickly identify and fix anomalous behaviors.

\begin{figure}[htbp]
  \centering
  \includegraphics[width=8cm]{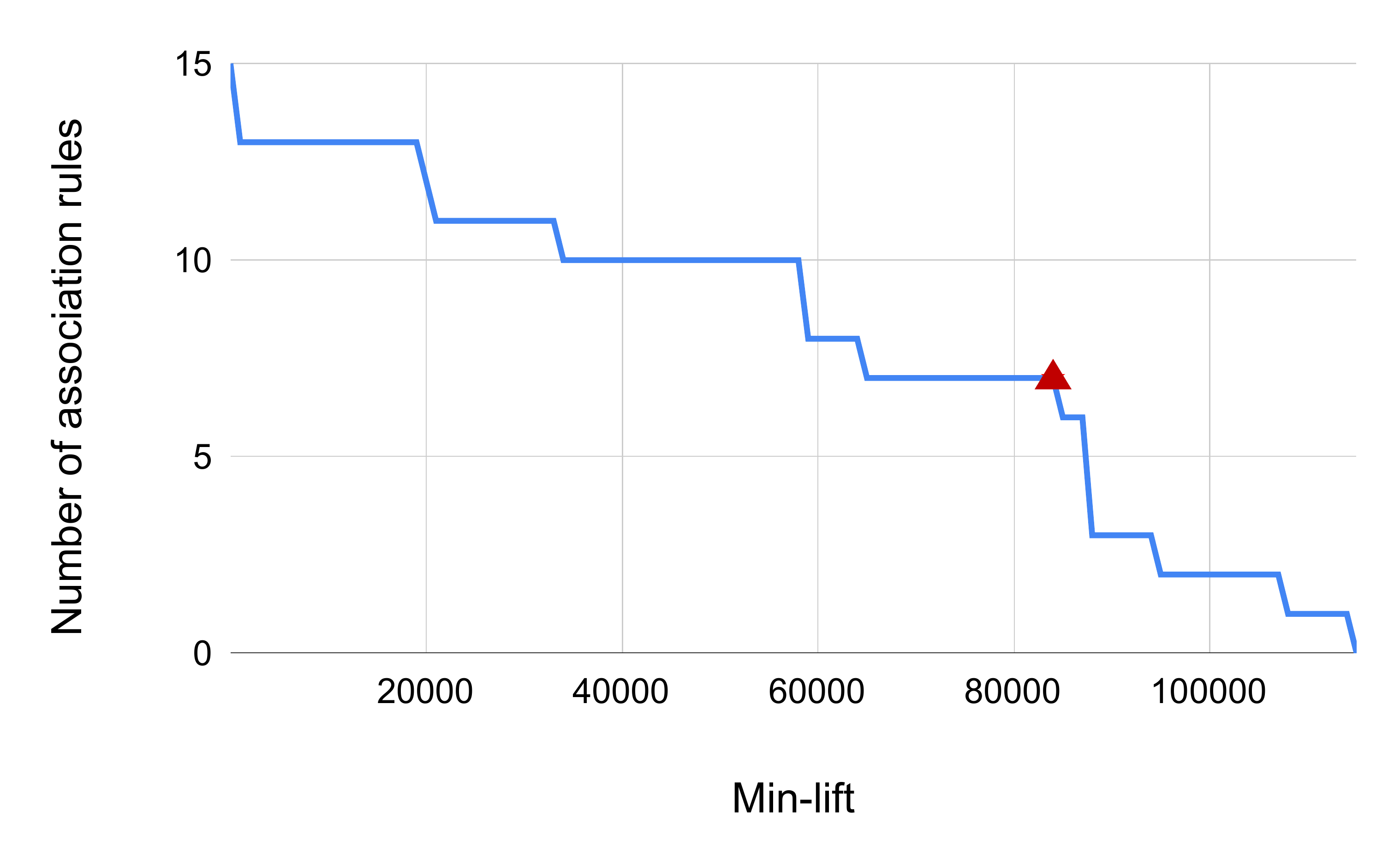}
  \caption{Number of association rules at different min-lift thresholds based on the SSH dataset. The triangle marks when a target rule appears in the use case.}
  \label{f:assoc_rule_vs_min_lift_ssh}
\end{figure}

Figure~\ref{f:assoc_rule_vs_min_lift_ssh} shows the number of association rules reported for different min-lift values. An actionable rule \textit{\{service, geographical location, SSH method\}} $\Rightarrow$ \textit{session failure} appears when min-lift becomes lower than approximately $88000$. Note that even though the support of this rule is only $0.1$, this rule is still very actionable because the dataset is complex and contains multiple types of failures at the same time. In other words, this example demonstrates how low-support rules can help us continuously improve the system as long as the lift is high, when our goal is not limited to investigating an urgent, major issue in the application.

\thispagestyle{plain} \section{Conclusions and Future Works}
\label{s:conclusion}

In this paper we have explored the problem of root cause analysis on structured logs and we have presented our scalable approach based on frequent item-set mining. We have also discussed a key change to \textit{support} and motivation for \textit{lift}, which are important frequent item-set metrics for our use case. Furthermore, we presented various optimizations to the core Apriori and FP-Growth frequent item-set algorithms including parallelism and pre- and post-processing methods. To our knowledge, this is the first work that utilizes frequent item-set paradigm at the scale of a large internet company for root cause analysis on structured logs.

In addition to the use cases described in Section~\ref{s:use_case}, we are exploring the use of the RCA framework on free-form text reports. To utilize RCA to surface features for different topics, the text reports first need to be labeled with topics. Both supervised \cite{zhang2015character} and unsupervised methods \cite{blei2003latent} can be applied for labeling the topics. An advantage of supervised model is that we can easily measure the quality of the inference and the topics are interpretable; however, it requires labeled data for training. Unsupervised approach does not require labeled data but the topics are often less interpretable, \textit{e.g.} each topic is often represented by top keywords~\cite{wang2007topical} and it is unclear how to measure the quality of the topic inference because there is no ground truth.

Given the topic labels, we can apply RCA on the text reports. As a result, RCA could detect significant features relevant to the labeled topics in the text corpus. For example, “reinforcement learning” and “speech recognition“ topics were extracted from a corpus of NIPS research papers \cite{wang2007topical} and potentially we can surface some features (\textit{e.g.} publish year) relevant to the topics. This is very useful for humans as it provides starting points for further investigation (\textit{e.g.} why are a set of features prevalent within a specific topic?).

\begin{figure}[htbp]
  \centering
  \includegraphics[width=5cm]{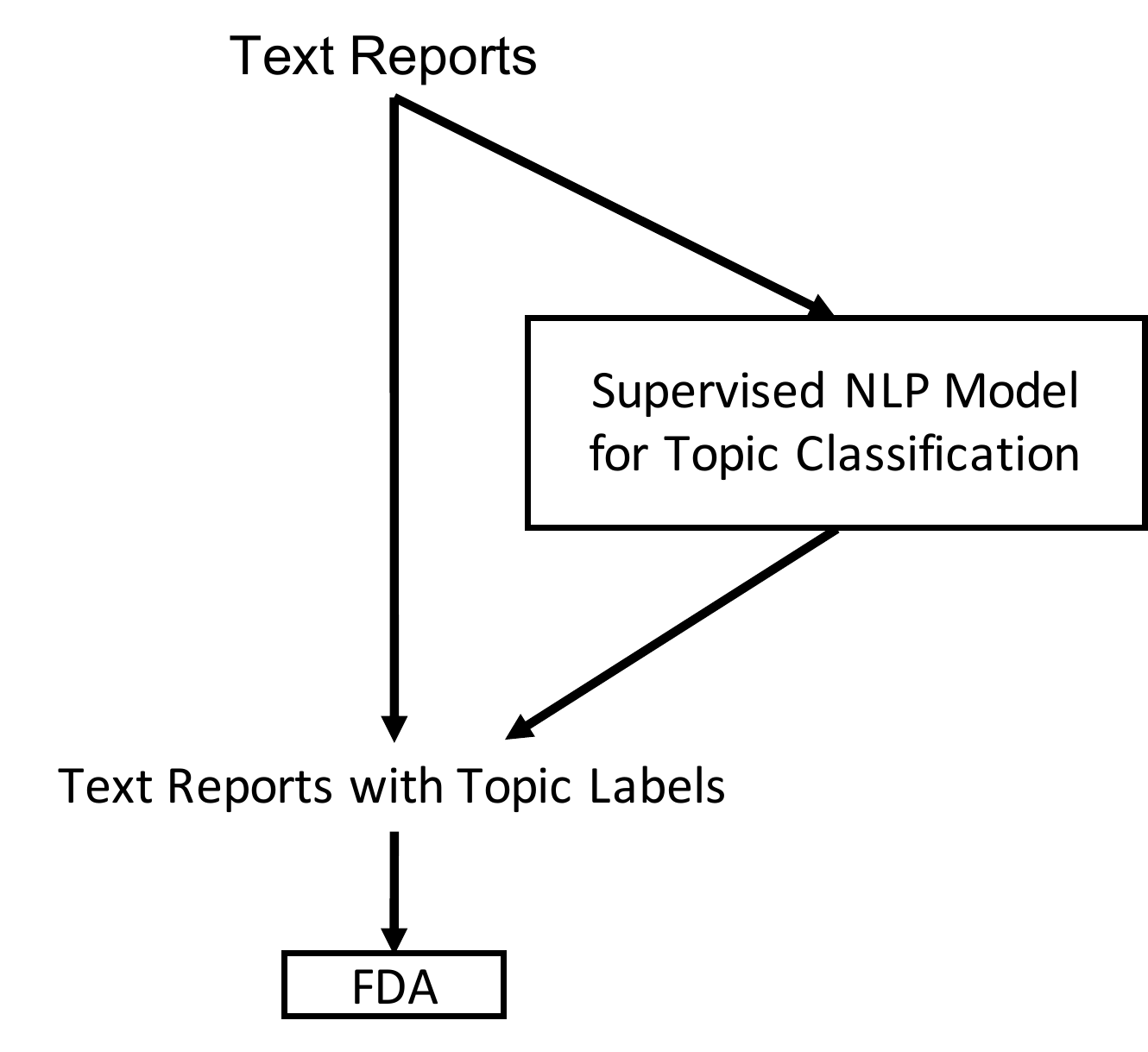}
  \caption{Flow of RCA on text-based reports.}
  \label{f:nlp_flow}
\end{figure}

As part of our future work we aim to focus on temporal analysis and on gathering more continuous feedback. Specifically, for temporal analysis, we are working on understanding the trend of association rules seen over time to discover seasonality and longer-term trends in order to catch degradation of services or hardware failures quicker. To overcome the lack of labels, we will focus on continuous feedback by leveraging our production system to understand which findings are more or less relevant to engineers in order to learn which findings to promote and which findings need to be hidden.






\bibliographystyle{ACM-Reference-Format}
\bibliography{myrefs}
\thispagestyle{plain} 
\end{document}